# Flexible and compact hybrid metasurfaces for enhanced ultra high field in vivo magnetic resonance imaging

*Rita Schmidt, Alexey Slobozhanyuk, Pavel Belov and Andrew Webb\**

R. Schmidt[1], Andrew Webb[1]
[1]Department of Radiology, Leiden University Medical Center, Leiden, Netherlands
E-mail: A.Webb@lumc.nl
A. Slobozhanyuk[2,3], P. Belov[2]
[2]Department of Nanophotonics and Metamaterials, ITMO University, St. Petersburg, Russia
[3]Nonlinear Physics Center, Australian National University, Canberra, ACT 2601, Australia

**Abstract**

**Developments in metamaterials and related structures such as metasurfaces have opened up new possibilities in designing materials and devices with unique properties. The main progress related to electromagnetic waves applications was done in optical and microwave spectra. Here we report about a new hybrid metasurface structure, comprising a two-dimensional metamaterial surface and a very high permittivity dielectric substrate that was designed to enhance the performance of an ultra-high field MRI scanner. This new flexible and compact resonant structure is the first one which can be integrated into a multi-element close-fitting receive coil array used for all clinical MRI. We successfully demonstrated the operation of the metasurface in acquiring vivo human brain images and spectra with enhanced local sensitivity on a commercial 7 Tesla system. These experimental findings prove the feasibility of real clinical applications of metasurfaces in MRI.**

Metamaterials offer a unique platform for controlling the propagation of acoustic and electromagnetic waves [1-4]. Developments in metamaterials [5-6] and related structures such as metasurfaces [7-10] have opened up new possibilities in designing materials and devices with unique properties [11,12]. Most applications of metamaterials have been demonstrated in the optical and microwave spectra using sub-millimeter sized unit cells. One of the potential clinical applications of metasurfaces is to magnetic resonance imaging (MRI), one of the most common diagnostic techniques used in hospitals worldwide, which operate at radiofrequency (RF) in the 63.5-300 megahertz range. Several studies have shown proof-of-principle implementations of metamaterials using lenses based on split rings [13], swiss-rolls [14[, discrete wires [15-17], magnetoinductive waveguides [18] and travelling-wave excitation [19]. However, since the size of the unit cells for metamaterials lies in the centimetre to tens-of-centimetre range, the vast majority of these implementations are based on three-dimensional metamaterial structures that have very large physical dimensions with respect to the available space within an MRI scanner. This is particularly problematic since all clinical MRI scanners use a large array of RF receive coils which are placed as close to the body as possible for maximum sensitivity. Large metamaterial structures mean that the coil array must be placed some distance away from the body, and the resulting loss in sensitivity cancels out the theoretical increase from the metamaterial. This means that there are currently no practical implementations of metamaterials in any clinical environment. In order to overcome this problem, thin and flexible metasurfaces [7-10] must be designed to be fully integrated into a multi-element receive array. Initial work has been performed by Algarin et al. using a thin (11 mm) metamaterial slab based on split-ring resonators [20], but this metamaterial introduced significant extra noise due to the large number of lumped elements, and does not have any flexibility in terms of spatially redistributing the near field magnetic and electric field components. In earlier work, a metasurface resonator was proposed in order to locally increase the sensitivity of 1.5 T MRI [16]. However, the use of such a structure in real practice is limited, due to the large physical size of the metasurface components which prevents them from being able to function with a dedicated receive array.

Our new approach is to design a thin, compact and flexible metasurface which can be placed between the patient and the close-fitting receive coil array. Specifically, we show applications of the hybrid metasurface, obtained by coupling a high permittivity pad with the metamaterial layers, in an examination of the human brain at 7 Tesla, concentrating on using the metasurface to produce a local increase in the SNR in the occipital cortex. Metasurfaces enable the control of electromagnetic waves in both linear [21] and nonlinear [22, 23] regimes. This is where artificial structures can be beneficial, since these can locally shape the RF field distribution in the region of interest. The results reported in this paper represents the first efficient practical metasurface-based device.

There are several different MRI-based approaches for achieving local increase in the SNR, which are complementary and can be combined with the metasurface approach. These include spatial control of the RF excitation by pulse design methods such as kT-points and spokes [24,25], as well as the use of multiple-transmit parallel excitation [26,27]. RF field shaping can also be achieved using high permittivity materials [28-30] or active off-resonant structures [31]. However, these two latter approaches lack the possibility of controlling the exact shape of the local field at subwavelength scales, which is possible using metasurfaces.

The new hybrid metasurface is shown schematically in **Figure 1**. Metallic strips with a thickness of a few tens of micrometres are attached to both sides of a flexible 8 mm thick pad made from a $CaTiO_3$ suspension in water with a relative permittivity of 110. The metasurface is designed to fit between the patient and the close-fitting RF receive coil array and to cover the region in the brain where an increase in MRI sensitivity is required (**Figure 1a**). The fundamental principle behind the enhancement phenomenon is the spatial redistribution and enhancement of electromagnetic near fields by the resonant excitation of the specific eigenmode of the metasurface. This phenomenon can be understood by considering the metamaterial's geometry (**Figure 1c**). The length of the long strips is designed to be slightly shorter than that required to produce the first half-wavelength resonance at 298 MHz. The specific eigenmode of the metasurface used for MRI has a similar magnetic field distribution to the $TE_{01}$ mode of a square dielectric resonator. It is interesting to note that an analogous structure has been employed to achieve a negative refractive index [32] in microwave applications. **Figure 1d** shows numerically calculated magnetic (left) and electric (right) field maps in the region of interest which is depicted as a dashed rectangle. Placing the metasurface near the patient modifies the field pattern of the RF coil due to the focusing of the magnetic field (**Figure 1d**) in the region-of-interest (ROI). In this way, the metasurface is able to provide higher local efficiency of the RF transmitted field and a higher image SNR due to higher receive sensitivity. It is important to note that, while in the case of negative refraction [32] the short wire pairs represent an artificial "magnetic atom" which displays magnetic resonance [33], in the current design we design the non-resonant strips in order to further modify the near field pattern of the eigenmode to obtain a greater increase in the MRI sensitivity and to perform fine tuning of the metasurface .

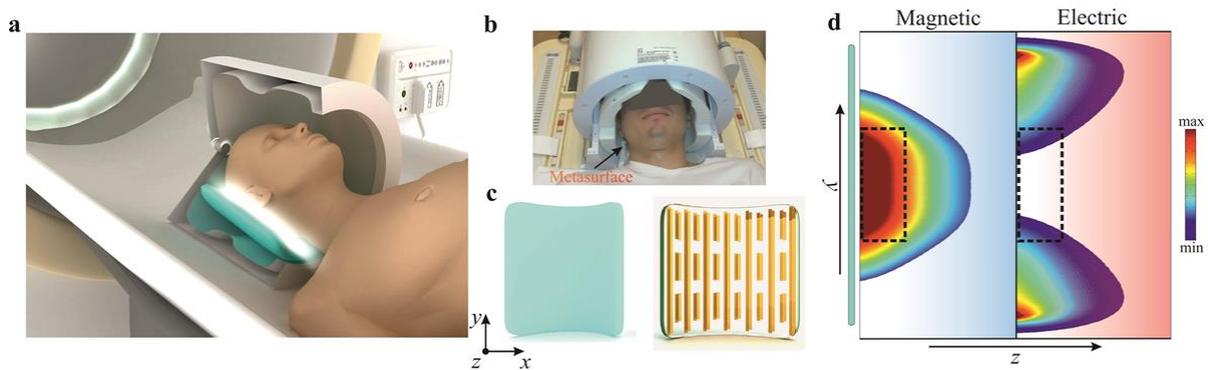

**Figure 1. Structural geometry of the metamaterial and simulation of near field magnetic and electric field distributions**. **a**, Schematic of the MRI setup with a cut-out for better visualization of the setup. **b,** A photograph of the in-vivo experiment including the transmit (outer) and multi-element receive coil array (inner). **c**, Artists view of the hybrid metasurface, including high permittivity dielectric substrate (left) combined with its metallic structure (right). d, Numerically calculated magnetic (left) and electric (right) field maps in vacuum near the metasurfaces (shown as a blue rectangle). The region of interest is depicted as black dashed rectangle.

**Results**

For the in-vivo experiment we were interested in the occipital cortex, which is commonly studied in anatomical, functional and spectroscopic studies of the visual cortex. Based on the results of electromagnetic simulations a metasurface resonating at 298 MHz was designed. Simulations and phantom experiments were performed to determine the performance of this metasurface. The phantom consisted of oil, since it has a very low relative permittivity of 5 and therefore there are no wavelength effects within the phantom at an operating frequency of 300 MHz. Three setups were used – the first without any structure in place, the second with an 8 mm thick CaTiO3 dielectric pad, and the third with the metasurface structure. The experiment was performed using a birdcage coil for both transmit and receive, and a low-tip angle excitation gradient echo sequence which produces images in which the SNR is proportional to the product of the RF transmit field ($B_1^+$) and the complex conjugate of the receive sensitivity ($B_1^{-*}$) divided by the square root of the accepted power which depends on the total resistance (see Methods for more details). **Figure 2** compares the electromagnetic simulations and the experimentally-acquired images. It can be seen in the simulation that the increase of the $B_1^+$ is approximately a factor-of-three close to the metasurface. Since the same RF coil is used to receive the signal, the $B_1^-$ also increases by a factor-of-three by the principle of reciprocity. In previous work an increase in the noise due to ohmic losses of a magnetoinductive lens based on split-rings has been reported [20]. However, in our experimental data we did not measure any increase in the noise level (calculated as the standard deviation of the noise in a region outside the phantom). This could be due to the fact that no lossy lumped elements are required, and also that the conductivity of human tissue at high fields is larger than at lower fields. We also measured the effect on the SNR of the metallic strips alone by replacing the high permittivity material with a plastic layer with permittivity of 1. These SNR measurements were repeated and no SNR change was observed due to the metallic strips.

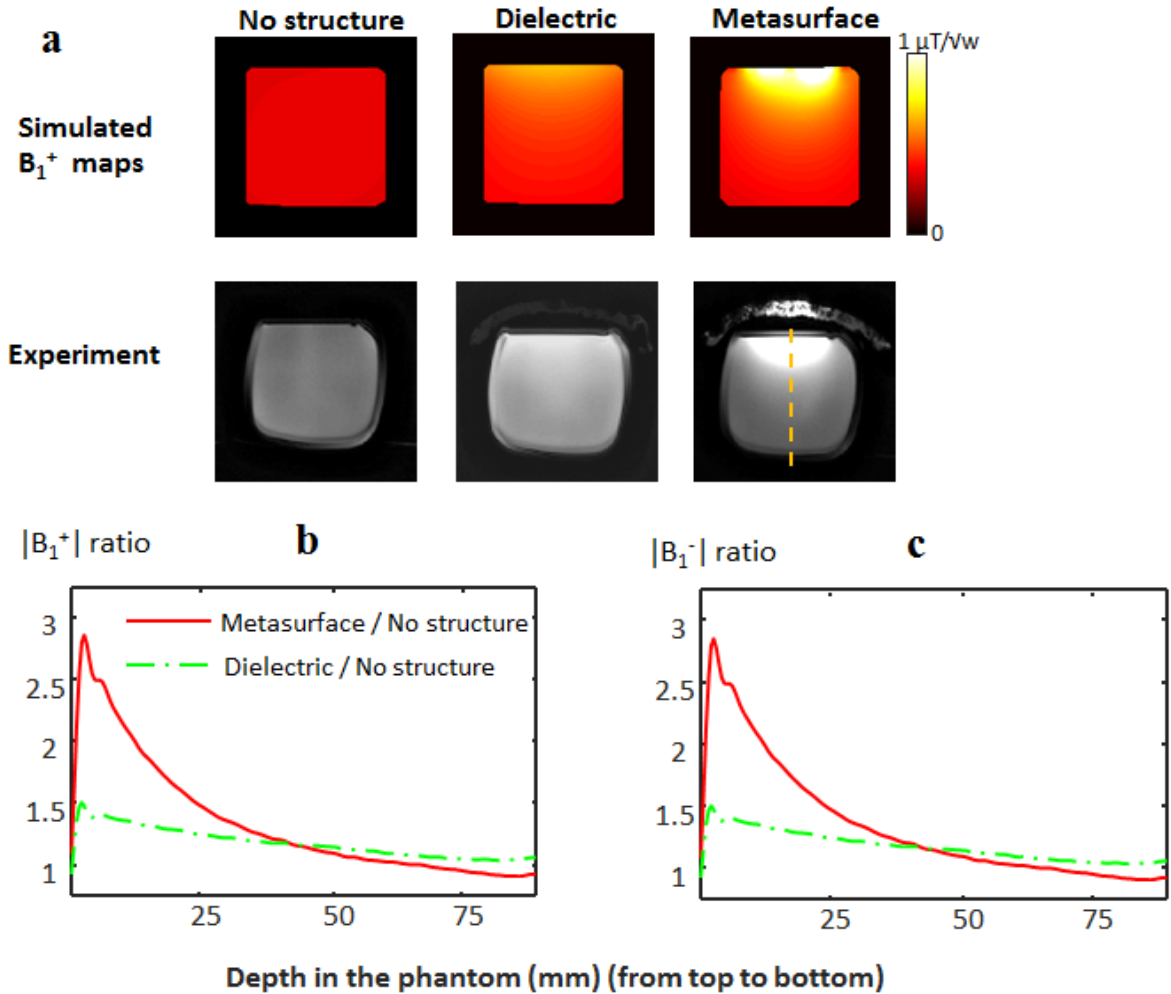

**Figure 2:** Experiments to simulate and measure the transmit ($B_1^+$) and receive ($B_1^-$) fields in an oil phantom. a, $B_1^+$ maps and the experimentally acquired low flip angle images: (left to right) no artificial structure, with dielectric material only and with the metasurface structure. b, Measured $B_1^+$ profile (along the yellow dashed line). c, Measured $B_1^-$ profile (along the yellow dashed line). The metasurface was placed on the top of the phantom.

**Figure 3** compares simulations of an electromagnetic model of the human head: (a) under normal scanning conditions, (b) with the tuned metasurface in place, (c) with the dielectric material only, and (d) with the copper strips only. The effect of the metasurface is clear, with a maximum enhancement ratio of approximately 2.7 compared to the normal scanning conditions.

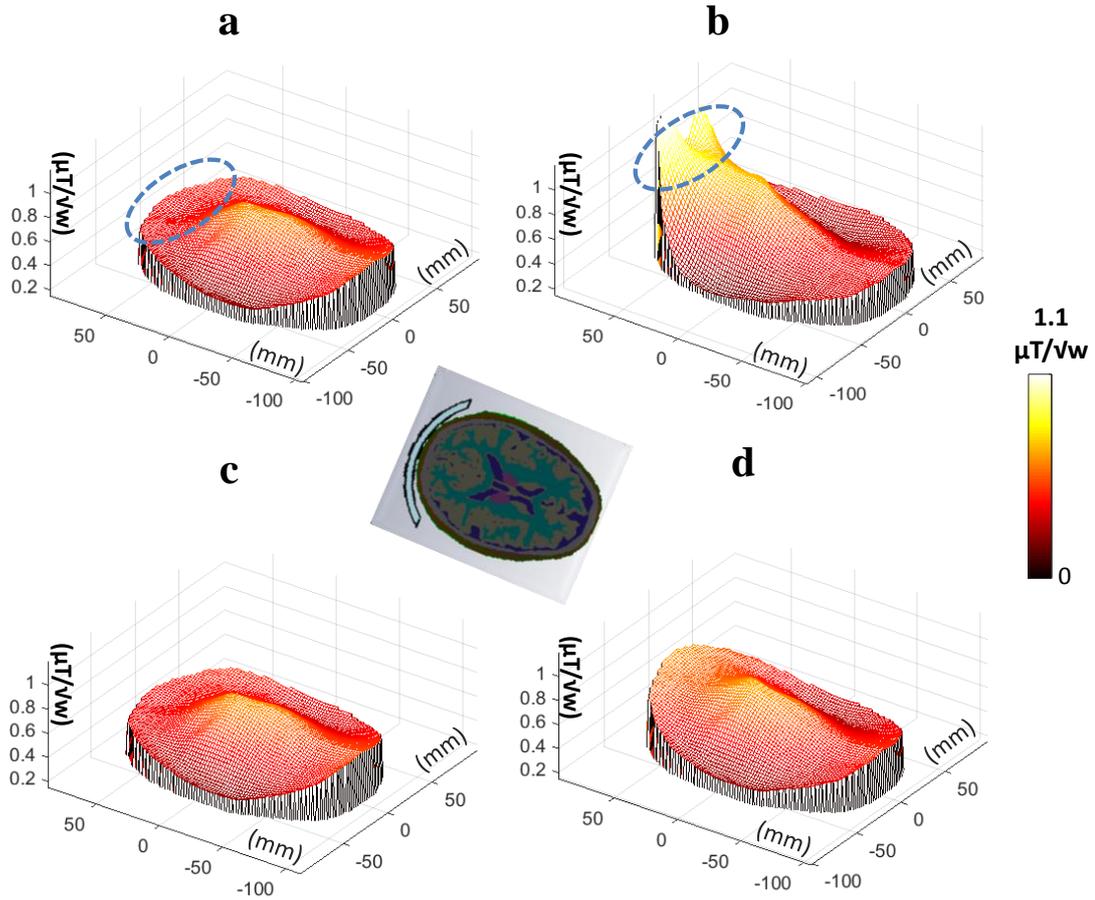

**Figure 3:** EM simulations of the full human model. a-d, Mesh plots of the $B_1^+$ distribution along a central axial cross-section of the brain for the following setups: (a) no structure, (b) with metasurface tuned at 7 T, (c) with metallic strips only, and (d) with the dielectric substrate only. The axial cross section of the brain and the metasurface location (in blue) is shown in the centre. The parameters of the resonant structure for this setup are given in detail in Methods. The blue dashed ellipse shows the region of interest in the occipital cortex.

In-vivo brain scanning was performed with the metasurface structure placed close to the occipital cortex, as shown in **Figure 1a**. A quadrature birdcage coil (Nova Medical NM-008A-7P) was used for RF transmission, and a close-fitting 32-channel array coil (Nova Medical NMSC-025-32-7P) for signal detection. We measured the effect of the metasurface on the isolation between different receiver elements in the 32-channel receive coil array, loaded with a phantom that simulated the human head, via $S_{12}$ measurements of different pairs of elements using a network analyzer. There were no systematic differences, with some channels showing slightly enhanced coupling (~1 dB) and others slightly reduced coupling (~1 dB). In addition to the magnetic field distribution, it is essential to consider the effect of the metasurface on the power deposition in the patient, which is expressed as the local specific absorption rate (SAR) averaged over 10 grams. Electromagnetic simulations show that the maximum local SAR in the brain was increased from 0.4 W/kg to 0.65 W/kg with the metasurface in place. This value lies well below the FDA limits for in vivo imaging.

**Figure 4** shows experimental maps (in one of the four volunteers studied) of the $B_1^+$ produced by the metasurface in the occipital cortex, derived from a low tip angle gradient echo image which is also displayed. Four volunteers were scanned and the average enhancement ratio for the RF transmit field was 2.0±0.3. Taking into account the increase in maximum SAR, the increase in transmit efficiency per square root of maximum SAR, the standard metric used in assessing RF coil efficiency, was a factor of 1.6. In practical terms this means that the amplitude or duration of the transmitted RF pulse can be reduced by a factor equal to the enhancement ratio of the RF transmit field.

The enhancement in the receive field was 1.9±0.2 (the enhancement in the receive field was normalized by the standard deviation of the noise in the image). The image SNR increases by a factor proportional to $B_1^-$ and in SNR-limited applications this increase can be converted into higher imaging resolution or reduced scanning times. An example of acquiring a localized $^1$H spectrum from a small area in the occipital lobe is demonstrated in **Figure 4c**. Compared to having no metasurface, an increase of 50% in the SNR of the spectra was obtained, which agrees well with the increase in the simulated $B_1^-$ field integrated over the spectroscopic volume. This corresponds to a decrease in total experimental time of a factor-of-two for constant SNR, which represents an important improvement for clinical applications of spectroscopy.

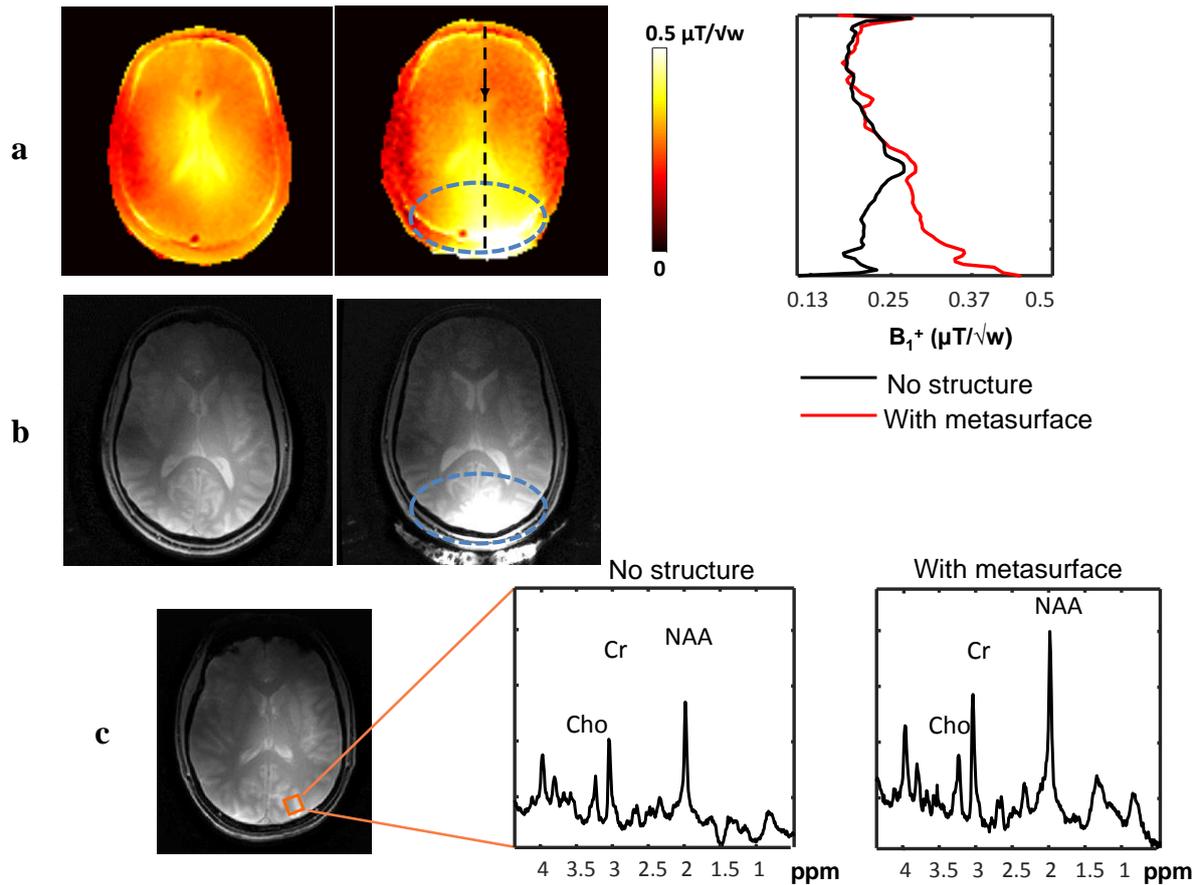

**Figure 4:** In-vivo imaging and spectroscopy results with and without the metasurface. (a) Measured RF transmit field ($B_1^+$) maps and $B_1^+$ profile along the dashed black line. (b) Low flip angle MR images of the brain. (c) Localized $^1$H spectroscopy performed without (left) and with (right) metasurface (the location of the voxel is shown by the orange overlay in the anatomic image).

**Discussion**

The metasurface described previously has been specifically designed for the maximum increase in transmit and receive sensitivity. However, the geometry of the metasurface has enormous flexibility in terms of tailoring the ability to "tune" its properties. For example, in many clinical applications the metasurface may be used to enhance areas in which the transmit efficiency is only tens of percent lower than the average. In this case a lower enhancement factor would be desired, and this could be achieved by geometry of the conducting elements in the metasurface. **Figure 5** shows plots of the simulated enhancement as a function of the thickness and permittivity of the substrate, as well as the length of the short strips. Additionally, by employing the concept of nonlinear metamaterials and metasurfaces [34, 35, 36] fine tuning can be performed electronically.

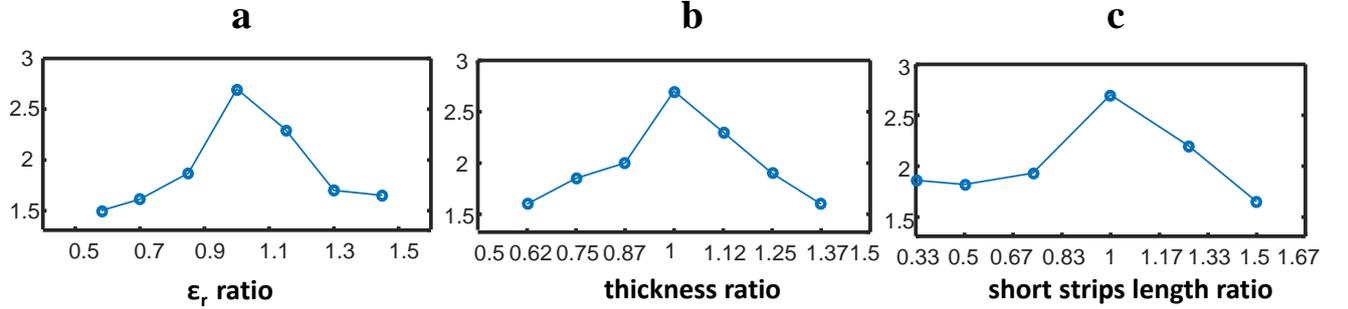

**Figure 5:** Parametric-dependence plots of the maximum enhancement ratio (averaged over a ROI of 6x6 mm) versus the examined parameter ratio to the value at resonance: (a) the permittivity ($\varepsilon_r$) of the dielectric substrate, (b) thickness of the dielectric substrate, and (c) the length of the short strips length.

In conclusion, this study has demonstrated a new design of a flexible and compact hybrid metasurface structure which can enhance the local RF transmit and receive efficiency in MRI. Since the design is thin and flexible it can be shaped to the anatomy of the patient, which is essential for combined operation with high density receive arrays. The structures enable manipulation of the magnetic field distribution in the region of interest, demonstrating the first applications of metasurfaces based devises for in-vivo imaging and spectroscopy of the brain, concentrating on the occipital cortex as a region of interest. This represents the first practical demonstration of a metasurface integrated with the multi-element receive arrays which are used in all clinical scans. We have shown results at very high field, but this approach can also be used at 3 Tesla and 1.5 Tesla which are currently used for most clinical studies.

**Methods**

The structure was constructed from 25 micrometre thick copper strips. The setup included long strips of 17.5 cm and a 3x3 matrix of short strips (3 cm length). The distance between the strips was 1 cm. The full structure size was 17.5x17.5x0.9 cm$^3$ including a 0.8 cm thick dielectric layer and a plastic sheet on which copper strips were attached. The high permittivity dielectric layer consists of an aqueous suspension of calcium titanate in water, with a relative permittivity of 110 and conductivity of 0.09 S/m ( $CaTiO_3$ to water volume ratio of 3:1 v/v), which allowed a flexible structure to be formed. The particular example used in this study was a single pad placed behind the head to increase the efficiency and sensitivity in the visual cortex region of the brain.

3D EM simulations were performed using finite integration technique (FIT) software (CST Microwave Studio, Darmstadt, Germany). The RF transmit field is defined as the left circularly polarized transverse field $B_1^+ = (B_{1x}+B_{1y})/2$ and the receive field as $B_1^- = (B_{1x}-B_{1y})/2$. The magnitude of the RF transmit field defines the excitation tip angle θ applied to the spins in the excited volume, $θ = γB_1^+τ$ where γ is the gyromagnetic ratio and τ is the pulse duration. The signal-to-noise (SNR) of the image is proportional to $sin(θ)· B_1^{-*}/\sqrt{P}$ [20], where P is the accepted power of the coil. All RF transmit ($B_1^+$) maps were normalized to an accepted power of 1 Watt. The simulation setup included a 16-rung high pass quadrature birdcage coil (inner diameter 30 cm; rung length 18 cm), corresponding to the transmit coil used for experimental measurements. The coil was loaded with the Virtual family model "Ella" dataset which consists of 76 different tissues with assigned values of permittivity and conductivity [37]. The mesh resolution was 1.0 x 1.0 x 1.0 $mm^3$. The phantom setup simulations used a rectangular shape oil phantom with either a flat metasurface structure or a simple high permittivity pad placed on top. In the in vivo simulations, the metasurface structure was curved to best fit the shape of the head. To account for the actual structure of the metasurface, in addition to the high permittivity layer, the simulation also included an interface between the copper strips and the dielectric layer of 0.5 mm thickness with $ε_r$=1.

Phantom and in-vivo images of a volunteer were acquired on a Philips Achieva 7 T MRI system. All methods were carried out in accordance with LUMC center guidelines and regulations. In vivo images and spectroscopy of the brain were acquired from healthy volunteers after informed consent was obtained in accordance with the Ethics Board of the hospital. Phantom experiments were performed using a quadrature head birdcage coil (Nova Medical NM-008A-7P) for both transmit and receive. In-vivo brain imaging was acquired using a quadrature birdcage for RF transmit and close fitting 32-channel receive head coil (Nova Medical NMSC-025-32-7P). The images included a standard gradient-echo sequence that was used for SNR estimation and $B_1^+$ maps images were acquired using the DREAM [38] sequence. The following scan parameters were used for gradient echo sequence: FOV 24x24 cm$^2$, spatial resolution 1.5 x 1.5 x 5.0 mm$^3$, TR/TE 10/3.4 ms, flip angle 5 °; and for DREAM sequence: FOV 24x24 cm$^2$, spatial resolution 2.5 x 2.5 x 5.0 mm$^3$, TR/TE 3/1.7 ms, $B_1^+$ encoding tip angle 50° and imaging tip angle 10°. The localized $^1$H spectroscopy used STEAM sequence with TE of 12 ms, mixing time of 13 ms and TR of 3000 ms, 15x15x15 mm$^3$ voxel, 64 averages.

**Acknowledgements**
We are grateful to I. Ronen for help with in-vivo spectroscopy experiment. This work was supported by an ERC Advanced Grant and NWO Topsubside (AGW). The work of A.S. and P.B. has been partially supported by Russian Science Foundation (Grant 15-19-20054).

**Author contributions**
Theoretical analysis and experiments were carried out by R.S. All authors analyzed and discussed the results. The manuscript was written by R.S., A.W., A.S. and P.B. and the figures were prepared by R.S., A.S.

**Additional information**
Competing financial interests: The authors declare no competing financial interests.

**References**

1. Pendry, J. B. and Smith, D. R., Reversing light with negative refraction, Phys. Today 57, 37–43 (2004).

2. Sihvola, A. Metamaterials in electromagnetics, Metamaterials 1, 2 (2007).

3. Kaina, N., Lemoult, F., Fink, M. & Lerosey, G. Negative refractive index and acoustic superlens from multiple scattering in single negative metamaterials, Nature 525, 77–81 (2015).

4. Pendry, J. B., Shurig, D. & Smith D. R. Controlling electromagnetic fields. Science 312, 1780–1782 (2006).

5. Engheta, N. and Ziolkowski, R. Electromagnetic Metamaterials: Physics and Engineering Explorations (Hoboken, NJ: Wiley-IEEE Press, 2006).

6. Cai, W. & Shalaev, V. Optical Metamaterials: Fundamentals and Applications (Springer, 2009).

7. Holloway, C. L. et al. An overview of the theory and applications of metasurfaces: The two dimensional equivalents of metamaterials, IEEE Antennas Propagat. Mag. 54, 10–35 (2012).

8. Kildishev, A. V., Boltasseva, A. & Shalaev, V. M. Planar photonics with metasurfaces, Science 339, 1232009 (2013).

9. Yu, N. & Capasso, F. Flat optics with designer metasurfaces, Nature Mater. 13, 139–150 (2014).

10. Glybovski, S.B., Tretyakov, S.A., Belov, P.A., Kivshar, Y.S. and Simovski, C.R., Metasurfaces: From microwaves to visible, Physics Reports, doi:10.1016/j.physrep.2016.04.004 (2016).

11. Zheludev, N. I. & Kivshar, Y. S. From metamaterials to metadevices, Nature Mater. 11, 917–924 (2012).

12. Fleury R, Monticone F and Alù A 2015 Invisibility and cloaking: Origins, present, and future perspectives, Phys. Rev. Appl. 4, 037001 (2015).

13. Freire, M. J., Jelinek, L., Marques, R. & Lapine, M. On the applications of μr = −1 metamaterial lenses for magnetic resonance imaging. J. Magn. Reson. 203, 81–90 (2010).

14. Wiltshire, M. C. K. et al. Microstructured magnetic materials for RF flux guides in magnetic resonance imaging. Science 291, 849–851 (2001).

15. Radu, X., Garray, D., Craeye, C. Toward a wire medium endoscope for MRI imaging. Metamaterials 3(2), 90-99 (2009).

16. Slobozhanyuk, A. P., Poddubny, A. N., Raaijmakers, A. J. E., van den Berg, C. A. T., et al.. A. Enhancement of Magnetic Resonance Imaging with Metasurfaces. Adv. Mater. doi: 10.1002/adma.201504270 (2016).

17. Jouvaud C., Abdeddaim R., Larrat B., and de Rosny J. Volume coil based on hybridized resonators for magnetic resonance imaging. Appl. Phys. Lett. 108, 023503 (2016).

18. Syms R. R. A., Young I. R., Ahmad M. M. and Rea M. Magnetic resonance imaging using linear magneto-inductive waveguides. J. Appl. Phys. 112, 114911 (2012).

19. Zivkovic, I., Scheffler, K. Metamaterial cell for B1+ field manipulation at 9.4T MRI. Proc. Intl. Soc. Mag. Reson. Med., 22, 4834 (2014).

20. Algarin J.M., Freire M.J., Breuer F., Behr V.C. Metamaterial magnetoinductive lens performance as a function of field strength. J. Magn. Reson. 247, 9-14 (2014).

21. Kapitanova, P. V. et al. Photonic spin Hall effect in hyperbolic metamaterials for polarization-controlled routing of subwavelength modes. Nat. Commun. 5, 3226 (2014).

22. Lee, J. et al. Giant nonlinear response from plasmonic metasurfaces coupled to intersubband transitions. Nature 511, 65–69 (2014).

23. Lapine, M., Shadrivov, I. V. and Kivshar, Y.S. Colloquium: Nonlinear metamaterials. Rev. Mod. Phys., 86(3), 1093-1123, (2014).

24. Cloos M.A., Boulant N., Luong M., Ferrand G., Giacomini E., et al.. kT-points: short three-dimensional tailored RF pulses for flip-angle homogenization over an extended volume. Magn. Reson. Med. 67, 72–80 (2012).

25. Sharma A., Bammer R., Stenger V.A., and Grissom W.A. Low Peak Power Multiband Spokes Pulses for B1+ Inhomogeneity-Compensated Simultaneous Multislice Excitation in High Field MRI, Magn. Reson. Med. 74, 747–755 (2015).

26. Katscher U., Börnert P., Leussler C., van den Brink JS. Transmit SENSE. Magn Reson Med. 49,144-50 (2003).

27. Grissom W., Yip C.Y., Zhang Z., Stenger V.A., Fessler J.A., and Noll D.C. Spatial domain method for the design of RF pulses in multicoil parallel excitation . Magn Reson Med.56,620-629 (2006).

28. Haines , K. , Smith , N.B. , Webb, A.G.. New high dielectric constant materials for tailoring the $B_1$ distribution at high magnetic fields. Journal of Magn. Reson., 203, 323-327 (2010).

29. Webb. A.G. Dielectric materials in Magnetic Resonance. Concepts Magn. Reson. Part A, 38, 148–184 (2011).

30. Yang, Q. X., Wang, J., Wang, J., Collins, C.M., et al.. Reducing SAR and Enhancing Cerebral Signal-to-Noise Ratio with High Permittivity Padding at 3 T. Magn. Res. Med. 65, 358–362 (2011).

31. Merkle, H., Murphy-Boesch, J., Gelderen, P. V., Wang, S., Li, T.-Q., et al.. Transmit $B_1$-field correction at 7T using actively tuned coupled inner elements. Magn. Reson. Med., 66,901–910 (2011).

32. Zhou, J., Zhang, L., Tuttle, G., Koschny,T. and Soukoulis, C.M. Negative index materials using simple short wire pairs. Phys. Rev. B, 73, 041101 (2006).

33. Lapine M. and Tretyakov S. Contemporary notes on metamaterials. *IET Microwaves, Antennas & Propagation*. 1(1), 3-11 (2007).

34. Lapine M., Shadrivov I.V., and Kivshar Y.S. Colloquium: Nonlinear metamaterials. Rev. Mod. Phys. 86, 1093 (2014).

35. Lee J. , Tymchenko M., Argyropoulos C., Chen P.Y., Lu F., et al. Nature. 511(7507), 65-69 (2014).

36. Minovich A. E., Miroshnichenko A. E., Bykov A. Y., et al., Functional and nonlinear optical metasurfaces. Laser & Photonics Reviews, 9, 195–213 (2015).

37. Christ, A., Kainz , W., Hahn , E.G., Honegger, K., Zefferer, M., Neufeld, E., et al.. The Virtual Family—development of surface-based anatomical models of two adults and two children for dosimetric simulations. Phys. Med. Biol., 55, N23–38 (2010).

38. Nehrke , K. , Börnert, P. DREAM—A Novel Approach for Robust, Ultrafast, Multislice B1 Mapping. Magn. Reson. Med. 68, 1517–1526 ( 2012).